\documentclass{article}
\usepackage{emulateapj}
\usepackage{graphicx}

\slugcomment{To Appear in ApJ}

\begin{document}


\title{Non-LTE Model Atmospheres for Late-Type Stars I. 
A Collection of Atomic Data for Light Neutral and Singly-Ionized Atoms.}


\author{Carlos Allende Prieto and David L. Lambert}
\affil{McDonald Observatory and Department of Astronomy, \\ 
The University of Texas at Austin, RLM 15.308, Austin, Texas 78712-1083}

\author{Ivan Hubeny\altaffilmark{1} and Thierry Lanz\altaffilmark{2}}
\affil{Laboratory for Astronomy and Solar Physics, \\
NASA Goddard Space Flight Center, Greenbelt, Maryland 20771}
\altaffiltext{1}{AURA/NOAO}
\altaffiltext{2}{Department of Astronomy,  University of Maryland,
College Park, MD 20742}

\begin{abstract}
With the goal of producing a reliable set of model atoms and
singly-ionized ions for use in building NLTE model atmospheres,
 we have combined  measured  energy levels, critically-compiled line
transition probabilities, and resonance-averaged calculations of
photoionization cross-sections. 
A majority of the elements from Li to Ca are considered, covering  
most of the important species in late-type atmospheres. These include
elements which contribute free electrons and/or continuous opacity in the
ultraviolet (e.g., Mg, and Si), as well as trace elements whose abundance
determinations rely on ultraviolet lines (e.g., B from B I lines).
The new data complement and, for the species in common,
supersede a previous  collection of model atoms  originally designed
for use in studies of early-type stars.  
\end{abstract}



\section{Introduction}

Classical -- Local Thermodynamical Equilibrium (LTE), plane-parallel,
horizontally homogeneous, hydrostatic equilibrium -- model stellar
atmospheres  have been providing a basic tool for stellar spectroscopy
for several decades, but are now known to suffer from various
weaknesses when applied to accurate observations of individual stars.
  Several lines of research have been adopted
to improve the models.  For example, in the context of late-type
stars, 
we mention here   three-dimensional hydrodynamical simulations 
of surface convection   (e.g. Stein \& Nordlund 1998,
Asplund et al. 2000);  improvements in molecular opacities (e.g.
Hauschildt et al. 1999a; Tsuji 2002); the adoption of spherical geometry  (e.g.
Plez, Brett, \& Nordlund 1992; Hauschildt et al. 1999b);  
and  semi-empirical modeling
(e.g. Allende Prieto et al.  2000). 

LTE  has been one of the strongest hypotheses invoked extensively  in
solving the coupled problems of radiative transfer and structure in
stellar atmospheres. This assumption  enormously  simplifies
the complexity of the calculations, since the populations of all ions and molecules
involved follow from the Boltzmann and Saha formulae. 
Yet, LTE has
been recognized to be inappropriate in many areas of the
Hertzsprung-Russell diagram.  Models that take 
departures from LTE into account are traditionally labeled non-LTE 
(or NLTE) models. The main area where the NLTE models have been
 applied are hot
stars (types O, B, A, hot white dwarfs and subdwarfs, and other 
objects with high-temperature atmospheres; see, e.g., Hubeny \& Lanz 1995).  
However, the advent of large telescopes with high-dispersion
 spectrographs, and the subsequent refinement of observations of late-type
 stars, call for an assessment of the possible departures from LTE on the
 model atmospheres for these stars.

Non-LTE calculations require detailed radiative and collisional rates
of all involved transitions in order to solve the statistical
equilibrium equations.  The role of photoionization cross-sections
should be emphasized, because changes of the degree of ionization  
 (with respect to the LTE ionization equilibrium) are a major 
non-LTE effect in stellar
atmospheres. Not all of the required energies and cross-sections for the
atomic/molecular processes that play a role in a stellar atmosphere are 
 available. Nonetheless, the  available data  has been
 expanded in recent years. Large collaborative projects,
such as the Opacity Project (see, e.g., Cunto et al. 1993),  the Iron
Project (see, e.g. Butler 1998), and SAM  (see, e.g. Brage, Judge, \&
Brekke 1996) have been enormously helpful.  
Long-term efforts on compilation of data, such as 
CHIANTI (e.g, Dere et al. 1997),  VALD (Kupka et al. 1999),  and  the Atomic
Spectroscopic Database (ASD) at the National Institute of Standards and
Technology (NIST),  suplemented by numerous contributions from  individuals and
small groups, are to be noted. 

The model atoms presented here are built for use in  the model stellar
atmosphere code {\sc Tlusty} (Hubeny 1988, Hubeny~\& Lanz 1995). A
family of model atoms and ions,  primarily in the context of hot
stellar atmospheres, has been made publicly available with the 
code\footnote{\tt http://tlusty.gsfc.nasa.gov/}.  
These models are briefly described
in \S 2. Section 3 is devoted to the 
new set of model atoms, describing the numerical
details of the smoothing process applied to the photoionization cross-sections, 
 and the construction of model atoms. 
 Some comments on the limitations and practicalities 
  for using these data, as well as a summary, are outlined in \S 4.

\section{The {\sc modion} Model Atoms}

In the context of  hot stellar atmospheres,  a collection of
model atoms for \ion{H}{1}, \ion{He}{1}, \ion{He}{2},
\ion{C}{1}--\ion{C}{4}, \ion{N}{1}--\ion{N}{5}, \ion{O}{2}--\ion{O}{6},
\ion{Ne}{2}--\ion{Ne}{4}, \ion{Mg}{1}--\ion{Mg}{2},
\ion{Si}{1}--\ion{Si}{4}, and \ion{S}{2}--\ion{S}{6} was prepared,
ranging from very simple to quite complete structures. These models are
essentially based on data extracted from TOPBASE. To manipulate these
large sets of atomic data, an interactive, IDL-based tool, was developed:
 {\sc modion} (Varosi et al. 1995; see also Lanz et al. 1996).
This interface program works with three files for each ion, containing
the level energies, the line oscillator strengths, and the
photoionization cross-sections, as extracted from TOPBASE. The energy
files have been updated with observed energies extracted from ASD/NIST 
(Martin 1997; Wiese 1997),
so that all transitions will be calculated at their actual frequency.

{\sc Modion} allows  model atoms of various sophistication to be built by
displaying a Grotrian diagram from which the explicit NLTE levels are 
interactively selected.
Although it might seem desirable to include as many  levels as possible,
this is often impracticable due to computing time and memory limitations.
To overcome these limitations, some individual levels are merged into
superlevels (see Hubeny~\& Lanz 1995). Low-excitation levels are generally considered as individual levels, while levels
of higher excitation are merged. The most excited levels are often not
treated explicitly, but are assumed to be in LTE with respect to the ground
state of the next ion. Model atoms and ions  have
been built for different purposes: simple model atoms (typically less than
15 explicit levels) are often quite appropriate for inclusion 
 in photospheric structure calculations, while very detailed models
(over 40 explicit levels) may be required for detailed line formation studies. 

After the level selection step, {\sc modion} builds a list of
bound-bound and bound-free transitions. Oscillator strengths from
TOPBASE are assigned to bound-bound transitions between individual
levels. {\sc Modion} takes care of the appropriate summing and averaging
for transitions between superlevels, supplementing TOPBASE data with
hydrogenic  values when data are missing. The necessary
sums are similarly performed  for photoionization cross-sections.
{\sc Tlusty} is able to deal with detailed and complex representations of the 
cross-sections,
 but in most cases we use approximate, yet adequate,  representations.  
Using {\sc modion}, we 
display the cross-sections in a $\log-\log$ plot of the photoionization cross-section vs. frequency,  and hand-pick some points with a cursor. 
The approximate cross-sections are
selected to be the lower envelopes of the detailed cross-sections, thus
the autoionization resonances are mostly neglected. Finally, when the
selections have been completed, {\sc modion} writes out a data file in the
format required by {\sc Tlusty}. Some limited upgrades have been made
in a later stage to incorporate Stark profiles for strong resonance
lines or to introduce fine-structure in strong resonance doublets like
 \ion{C}{4} $\lambda 1550$ or 
\ion{Si}{4}  $\lambda 1496$.

\section{The RAP Model Atoms}

This new set of model atoms is, like the {\sc modion} models, 
based on data from TOPBASE and NIST. They treat, however, the photoionization
cross-sections differently, and in more detail. As they are new and have 
never been used in calculations of radiative transfer and model atmospheres
before, we describe them in depth. These models are, at the time of this
writing, available for 
 neutral  Na and S,  singly ionized Ne, 
   and  neutral and singly-ionized  Li, Be, B, C, N, O, F,
Mg, Al, Si and Ca.

\subsection{Resonance-Averaged Photoionization Cross-Sections}

It has been acknowledged that errors in the computed energies of the
atomic levels may result in spurious shifts of the frequencies at which
strong photoionization resonances are predicted by 
theoretical calculations. Consequently, 
several authors have suggested different procedures to smooth the
 Opacity Project 
 data (see, e.g., Verner et al. 1996; Bautista, Romano, \& Pradhan 1998).
 In addition, smoothing the cross-sections reduces sometimes the number 
 of data points to a more manageable level.

Bautista et al (1998) suggest a recipe for a convenient Gaussian
smoothing.  They recommend a value for the width of the Gaussian of
$\sigma = 0.03 E$, where $E$ is the energy of the ionizing photon. They
apply such a smoothing to the photoionization cross-sections of
ground levels of all atoms and ions included in TOPBASE and to new data
on \ion{Fe}{1}$-$\ion{Fe}{5} provided by the Iron Project, producing
Resonance-Averaged Photoionization (RAP) cross-sections. While in
low-density environments, such as gaseous nebulae, virtually all
atoms and ions are in ground states, this is not the case for  stellar
atmospheres. Extending the calculations of RAP cross-sections to the
rest of the levels is therefore desirable. We
deal with neutral and singly-ionized species of the elements included
in TOPBASE that have been described by $LS$ coupling in ASD/NIST,
excluding H and Fe. In fact, the list covers most of the atomic species of
relevance for studying late-type stellar atmospheres.

We have closely followed the procedure described by Bautista et al. (1998).
The RAP cross-section at a given energy, $\sigma_A(E)$, is computed 
from the TOPBASE photoionization cross-section, $\sigma(E)$, 
through the  integral

\begin{equation}
	\sigma_A (E) = \frac{\displaystyle \int_{E_0}^{\infty} {\displaystyle \sigma (x) \exp \left[ \frac{-(x-E)^2}{2 (\delta E)^2} \right] }dx}{\displaystyle \int_{E_0}^{\infty} {\displaystyle  \exp \left[ \frac{-(x-E)^2}{2 (\delta E)^2} \right]}   dx}
\end{equation}

\noindent where $E_0$ is the ionization threshold energy, and 
$\delta E = 0.03 E$.  $E_0$ is 
slightly lower than  the energy difference between the level and the
continuum, due to line merging near the series limit.

\begin{figure*}[t!]
\begin{center}
\includegraphics[width=6cm,angle=90]{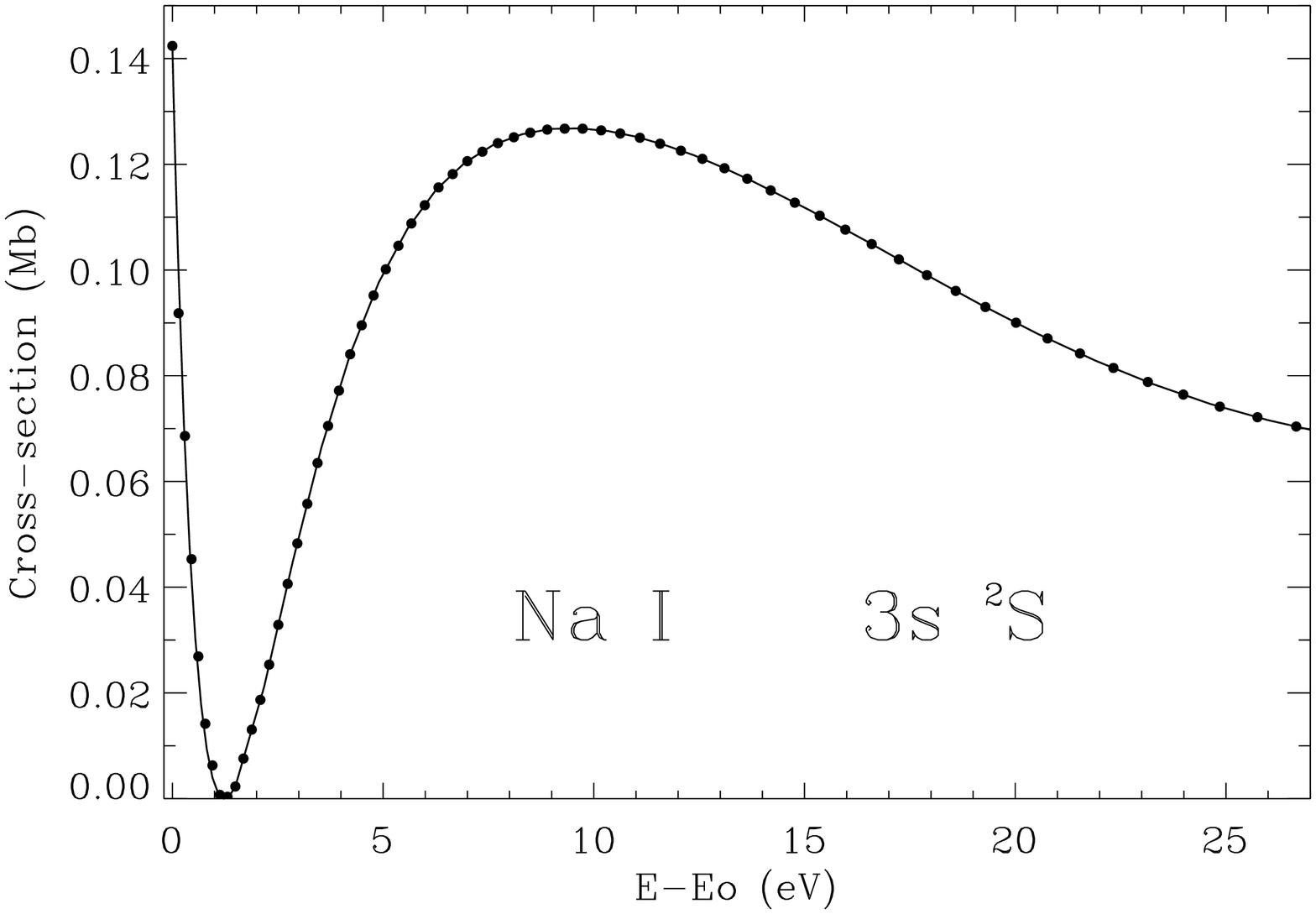}
\includegraphics[width=6cm,angle=90]{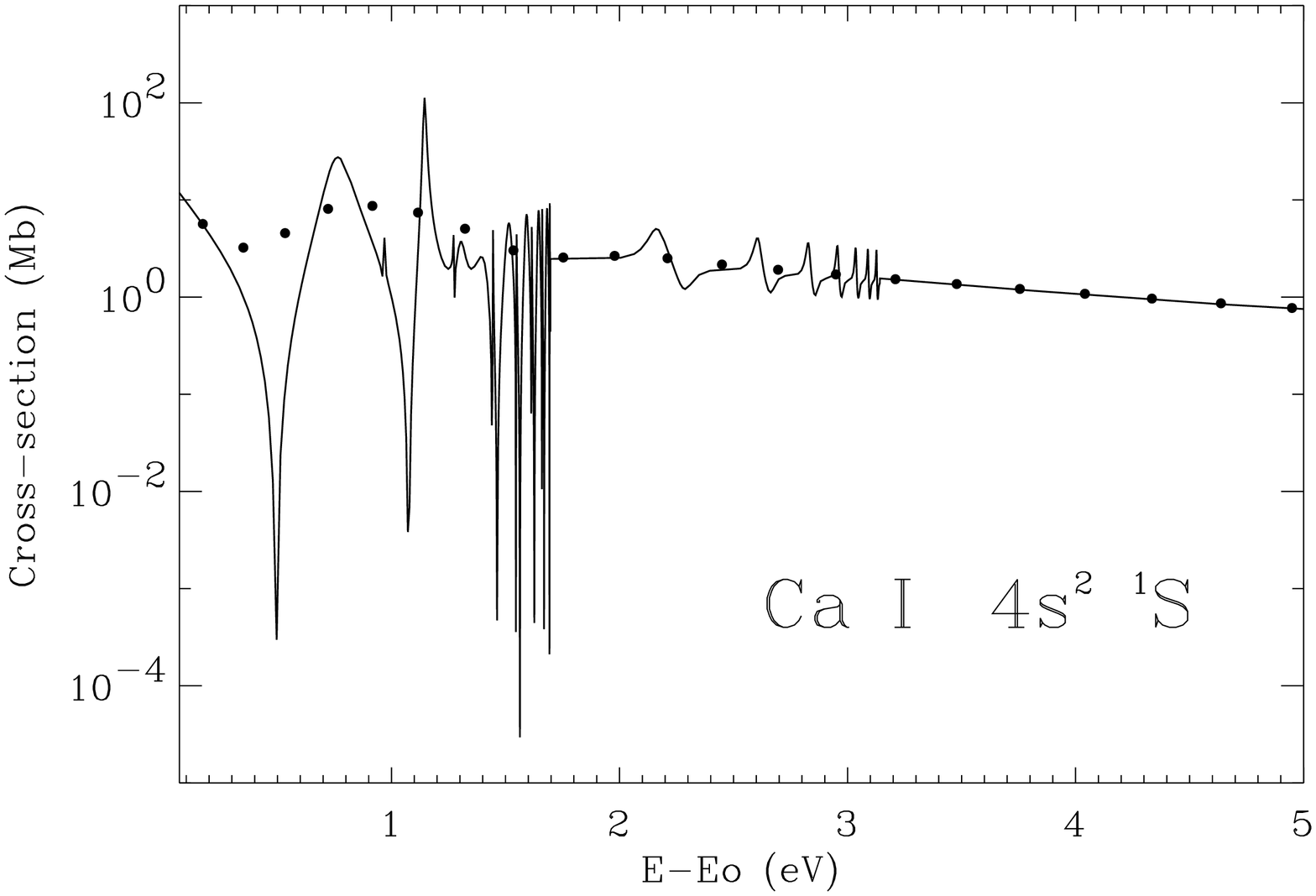}
\figcaption{
Detailed (solid line) and resonance-averaged (filled circles) 
cross-section for the ground states of \ion{Na}{1} and \ion{Ca}{1}. These
two cases exemplify a smooth cross-section and a complex resonance structure. 
\label{fig1}}
\end{center}
\end{figure*}

To speed up  the calculations, without losing accuracy, we restricted the 
limits of the integral (1) to $E \pm 5  \delta E$. 
In order to avoid unwanted systematic effects at the minimum (threshold) 
and maximum energies in the TOPBASE calculations, we performed a linear 
interpolation over the energies for which less than five points were 
available within $\pm 5  \delta E$.  
The RAP cross-sections were derived at energies 
separated by the smallest steps allowed by the sampling theorem, 
using the recurrence formula $E(i+1) = E(i) (1 + \delta E/E)$ from 
$E =  E_0$ to $E \le E_{MAX}$. Figure 1 shows two  examples 
corresponding to the ground levels of \ion{Na}{1}, and \ion{Ca}{1}:  
the thin solid line corresponds to the original cross-sections, 
and  the filled dots represent the RAP cross-sections. 

\subsection{Assembling the RAP Model Atoms}

The following sections describe how radiative and collisional processes
are accounted for in the the RAP model atoms. Only collisions with electrons
are included. Collisions with neutral H atoms may be important in
late-type stars, but lacking  term-dependent or  species-dependent
data, they are not considered in these models. The phenomenological
approximations usually found in the literature are best implemented directly
in the NLTE codes, rather than in the input data files.

\subsubsection{Radiative processes}

The Opacity Project provides energies and oscillator strengths  for
atomic  levels and radiative transitions among them. However, even though
this database provides a large amount of atomic data of sufficient
quality  for many purposes, theoretical
calculations cannot, with few exceptions, predict the energies of 
the levels with accuracy. As a result, the 
wavelengths of the lines cannot be predicted 
to the precision available to laboratory  spectroscopists. 
There are also measurements of transition probabilities that may be more 
accurate than those in the OP database.

We have chosen to use the atomic energy levels and the spectral lines
included in ASD/NIST, version 2.0, to build
models for the same atoms and singly-ionized ions for which the have
derived RAP cross-sections from TOPBASE data. The photoionization 
cross-sections
are given as a function of the energy from the threshold, and therefore they
are simply shifted to the observed energy levels.
TOPBASE ignores fine
structure, and so have we,  by grouping the split levels into a
single one with an averaged energy, using $(2{\rm J}+1)$ as weights
(where J is the quantum number that represents the total angular
momentum),  according to  the relative populations of the levels
expected in thermodynamic equilibrium. Transitions between
fine-structure states of a given level are ignored.

A scaled hydrogenic photoionization cross-section  has been assigned to
the levels whose  photoionization cross-sections were not listed in
TOPBASE, assuming that the ionization takes place to the ground state
of the next ion:

\begin{eqnarray}
\begin{tabular}{ccc}
$\sigma(\nu)$   = & $2.815 \times 10^{29} \displaystyle\frac{(Z+1)^4}{\nu^{3} n^{5}} 
g_{II}\left(n,\displaystyle\frac{\nu}{(Z+1)^2}\right)$ & ${\rm cm}^{2}$  \\
 & & \\
 &  $(\nu \ge \nu_0 \equiv E_0/h)$, &  \\
\end{tabular}
\end{eqnarray}

\noindent  where   $h$ is Planck's constant, $n$ is the quantum
principal number of the last electron in the  configuration, 
$\nu$ the frequency, $Z$ the charge of the ion (i.e.,
$Z=0$ for neutrals, $Z=1$ for once ionized, etc.),  and $g_{II}$ the
Gaunt factor expressed as

\begin{equation}
g_{II} (n,\nu)= \sum_{i=1}^{7} C^n_i \nu^{i-4}
\end{equation}

\noindent  with the coefficients $C^n_i$ given by 
Mihalas, Heasley, \& Auer (1975). C.g.s. units are used throughout  the paper.

The radiative transitions between fine-structure states ($i$,
$j$) of two levels ($l$ and $u$) have been summed to a single 
one with a $gf$-value:

\begin{equation}
g_l f_{lu}  =  \sum_{i} g_i \sum_{j} f_{ij} 
\end{equation}

\noindent where $g_i=2{\rm J} + 1$ represents the statistical 
weight of a fine structure  level. 

Allowed and forbidden radiative transitions whose
$f$-values are listed in  the NIST database have been included in the
models\footnote{We note that, in some particular cases, NIST may list 
theoretical calculations less accurate than OP.}.  Also, allowed radiative 
transitions not listed in NIST
have been included, assuming a scaled hydrogenic transition probability

\begin{equation}
f_{ij} = f_{ij}^{\rm H} \frac{g_i}{2n_i^2}.
\end{equation}

When more than one level in TOPBASE -- with separate photoionization
cross-sections -- was listed  with identical energies in
the NIST lists, they  were included in the models with
 their respective RAP cross-sections. The states described
in the NIST database with a coupling other than $LS$ have  not been
included,  but states with configurations, terms, or energies listed as
uncertain are included when matched by the TOPBASE states within 0.08
Ryd. Only levels with energies lower than the series limit are kept. 
As an example, Figure 2 displays the Grotrian diagrams for
\ion{Na}{1} and \ion{Si}{1}. Table 1 lists the number of levels and
radiative transitions considered for each model.

\begin{figure*}[t!]
\begin{center}
\includegraphics[width=6cm,angle=90]{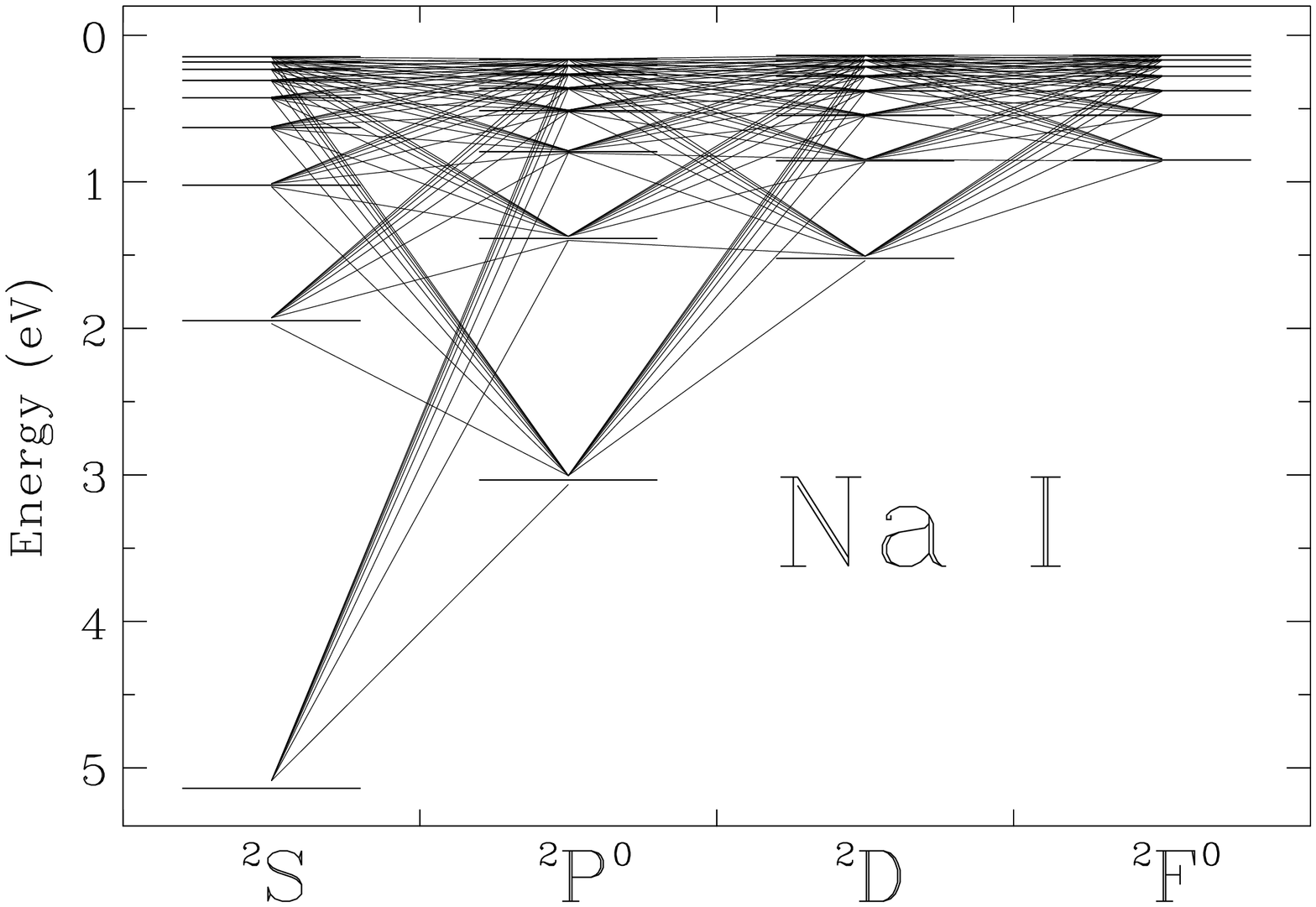}
\includegraphics[width=6cm,angle=90]{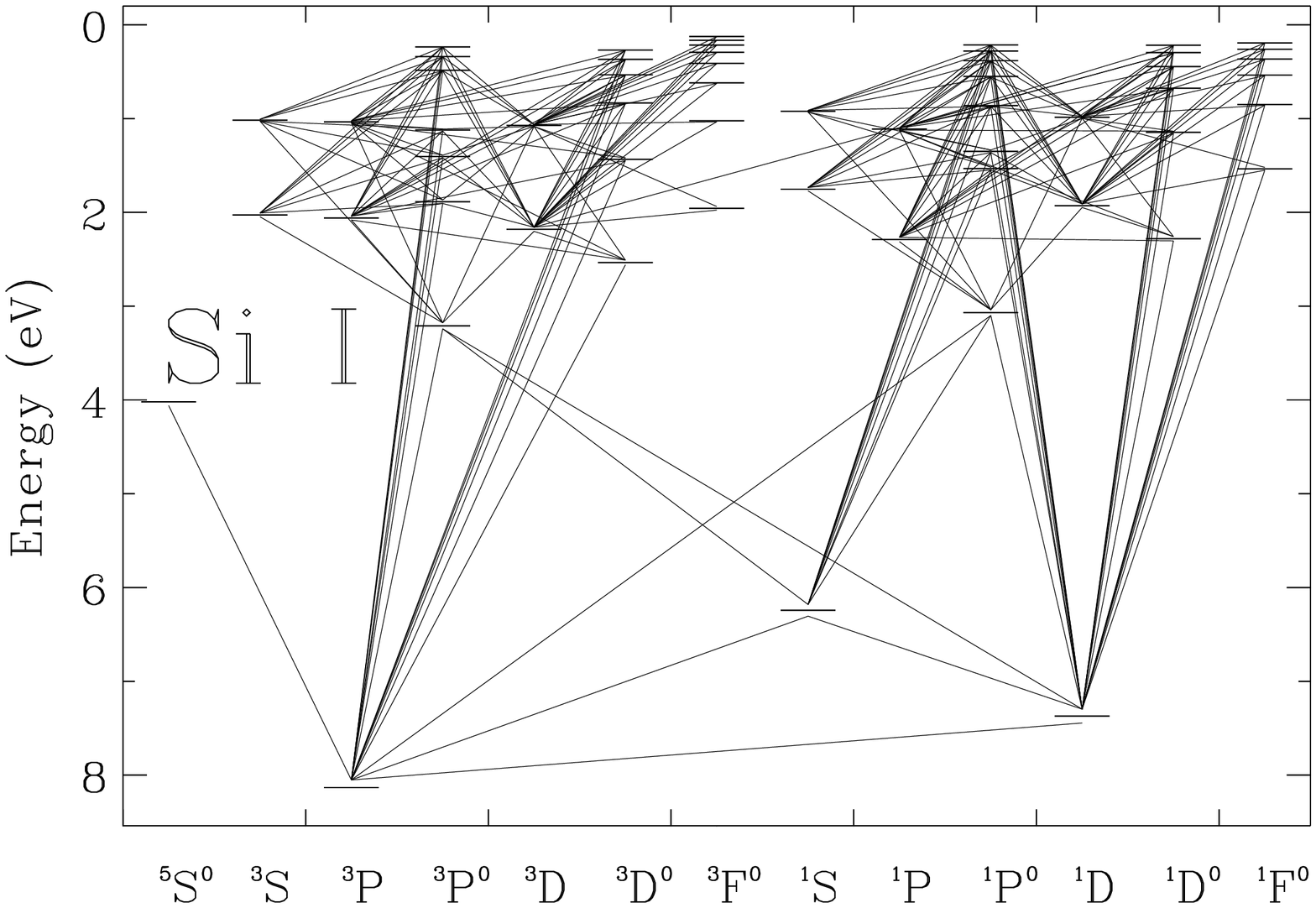}
\figcaption{
Energy levels for  \ion{Na}{1} and \ion{Si}{1}. \label{fig2}
}
\end{center}
\end{figure*}

We have formatted the described data  to be read by the program {\sc
Tlusty}.  The code is designed to build non-LTE model atmospheres, and
it will be the tool we shall employ in future analyses of stellar
spectra.  There are several documents explaining the data format in
detail (Hubeny 1988; Hubeny \& Lanz 1997), hence we shall not describe
it here. In the files, a series of  {\it options} that apply to the
solution of the statistical equilibrium  
equations have been chosen. These options set 
several input parameters (e.g. collisional ionization and excitation,
depth-dependency of the line absorption profiles, etc.), which are
briefly described below, but they can be changed at ease  by the
user (see Hubeny 1988). Free-free absorption for metals is accounted for by
a hydrogenic cross-section with the Gaunt factor set to one, 
while the exact Gaunt factors are used for H and He II. 
The line absorption profiles are assumed Gaussian and
depth-independent.  All  levels  are connected with each other
 and with the continuum by collisions with electrons. How these collisional
 processes are treated is described in more detail below. Mostly, but not
 all, applies as well to the {\sc modion} models.
 
 \subsubsection{Collisional processes involving electrons}
 
The collisional rate for the transition $i \rightarrow j$ 
is given by $n_e q_{ij}$,  where $n_e$ is the number density of
electrons and $q_{ij}$  is the excitation rate coefficient. Excitations
induced by electron collisions between states connected by 
radiatively permitted  transitions  are accounted for using  
Van Regemorter's formula (Van
Regemorter 1962, Mihalas 1972), with the excitation rate 
coefficient
as:

\begin{eqnarray}
\begin{tabular}{ccc}
$q_{ij}$ &  =              & $\pi a_0^2 \sqrt{\frac{\displaystyle 8k}{\displaystyle  m_e \pi}} 
\frac{\displaystyle 8 \pi}{\displaystyle \sqrt{3}} \sqrt{T}  f_{ij}  \left( \frac{I_{H}}{E_{ij}} \right)^2 U_0 e^{-U_0} \Gamma(U_{0})$ \\
	   &		&														\\			
           & $\simeq$   & $19.7  \frac{\displaystyle f_{ij}}{\displaystyle T^{3/2} U_0} e^{-U_0} \Gamma(U_0) ~~{\rm cm}^3 {\rm s}^{-1}$, \\ 
\end{tabular}
\end{eqnarray}

\noindent where $a_0$ is the Bohr radius, $k$ is Boltzmann's constant, 
$m_e$ is the mass of the electron,
$I_H$ is the threshold ionization energy for hydrogen, 
$T$ is the electron temperature, $U_0= E_{ij}/kT$,  and for ions

\begin{equation}
\Gamma (U_0) = max \left[\overline{g}, \frac{\sqrt{3}}{2\pi} e^{U_0}E_1(U_0)\right]
\end{equation}

\noindent  with $\overline{g} = 0.7$ (transitions between levels with the
same principal quantum number) or $\overline{g}= 0.2$ (otherwise) 
 (Mihalas 1978), $E_1$ is the first-order exponential integral, 
 but for neutral atoms

\begin{eqnarray}
\begin{tabular}{cccc}
$\Gamma (U_0)$ & =  & $\frac{\displaystyle \sqrt{3}}{\displaystyle 2\pi}  e^{U_0}E_1(U_0)$, &  $U_0 \le 14$ \\
		&	&								    &		\\
$\Gamma (U_0)$ & =  & $\frac{\displaystyle 0.066}{\displaystyle \sqrt{U_0}} \left(1+ \frac{\displaystyle 3}{\displaystyle 2U_0}\right)$, &  $U_0 > 14$ \\
\end{tabular}
\end{eqnarray}

\noindent (Auer \& Mihalas 1973). Collisional excitation between levels 
linked by forbidden radiative transitions is considered by means of
 the Eissner-Seaton formula (Seaton 1962):

\begin{equation}
q_{ij} = \frac{8.631 \times 10^{-6}}{g_i \sqrt{T}} e^{-U_0} \Upsilon_{ij}
 ~~{\rm cm}^3 {\rm s}^{-1}, 
\end{equation}

\noindent    adopting, $\Upsilon_{ij}=0.05$ for
all cases.  Pradhan \&  Peng (1995) and Pradhan \& Zhang (2000) review some of
the detailed calculations and scaling laws of effective collisional strengths 
that are available for particular ions. Those data 
should update the approximate values adopted here.   

\begin{figure*}[t!]
\begin{center}
\includegraphics[width=6cm,angle=90]{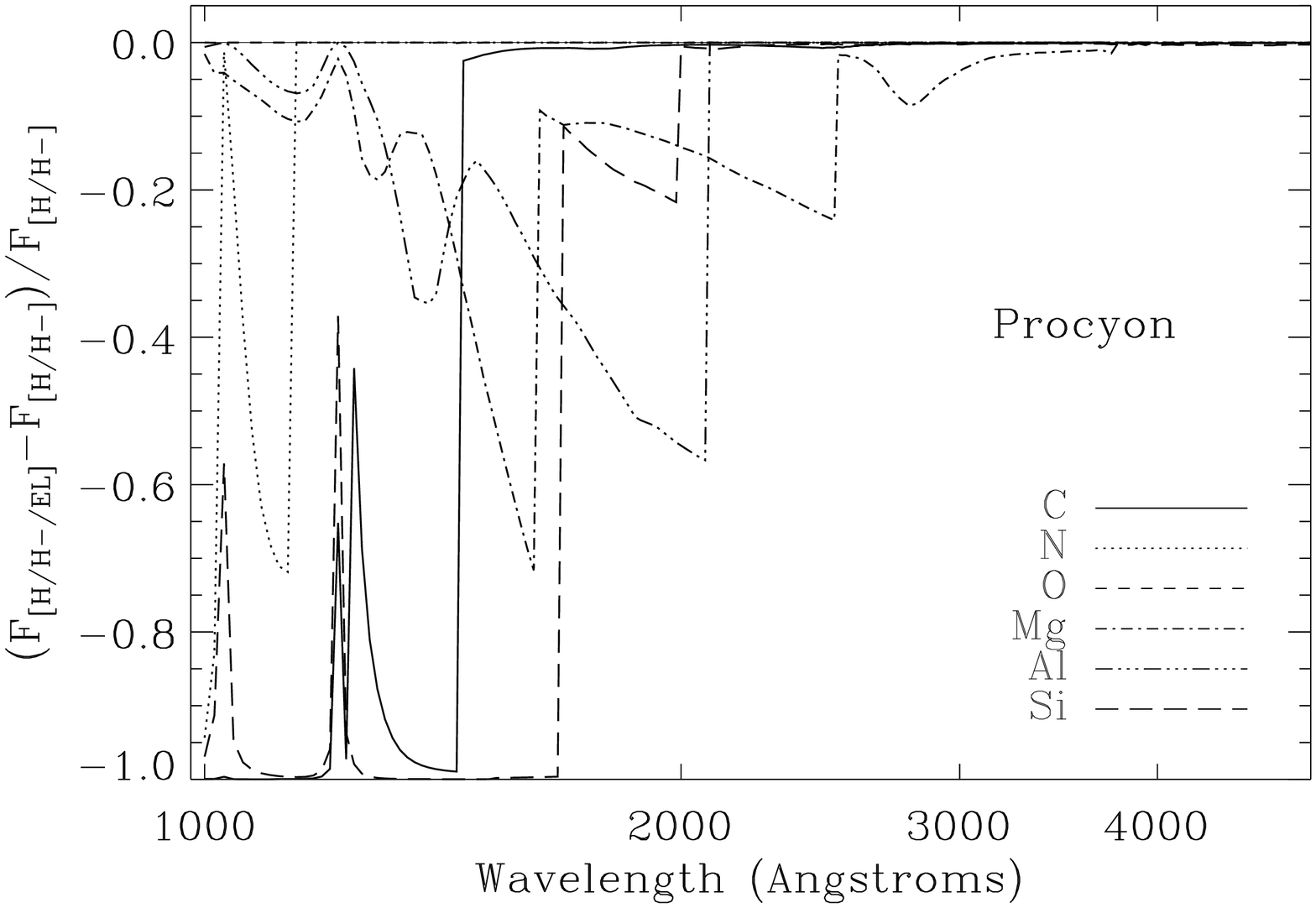}
\includegraphics[width=6cm,angle=90]{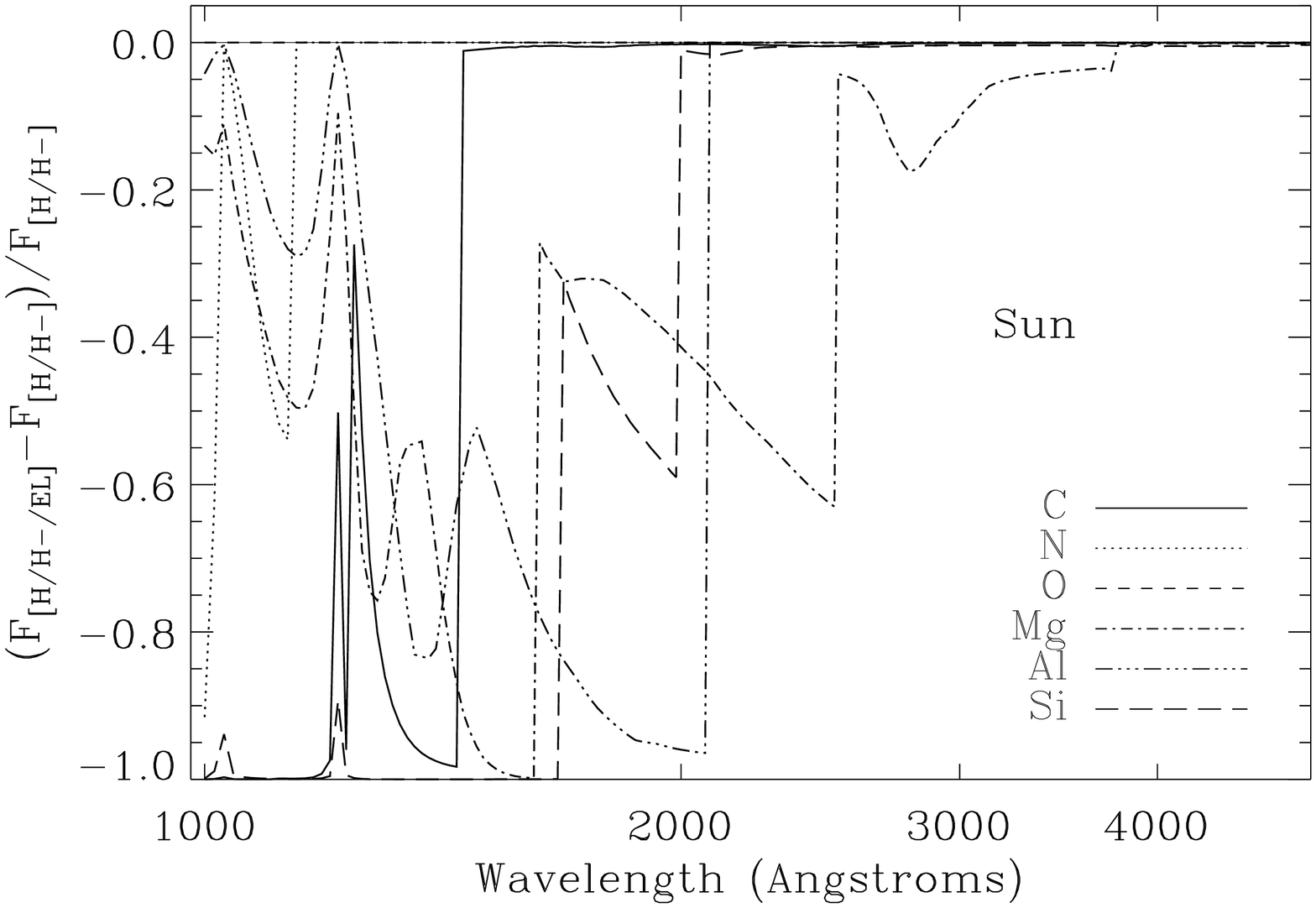}
\figcaption{
Relative difference between the bound-free  continuum flux 
from H and H$^{-}$,  and the same plus the contribution 
from C, N, O, Mg, Al, or Si, for
Kurucz's model atmospheres with the stellar parameters of the Sun (G2V) 
and Procyon (F2 IV-V). \label{fig3}
}
\end{center}
\end{figure*}

We also used an approximate formula (Seaton 1962) 
to evaluate  collisional ionizations:

\begin{equation}
q_{ij} = \frac{1.55 \times 10^{13}}{U_0 \sqrt{T}} e^{-U_0} \sigma_0 ~~{\rm cm}^3 {\rm s}^{-1},
\end{equation}

\noindent  with the photo-ionization cross-section at 
the threshold ($\sigma_0$)  
from  TOPBASE cross-sections (when available), 
or the hydrogenic approximation (with the Gaunt factor equal to one)

\begin{equation}
\sigma_0 = 7.91 \times 10^{-18} \frac{n}{(Z+1)^2} ~~{\rm cm}^2
\end{equation}

\noindent where  $n$ is again the principal quantum number 
of the last electron and $Z$ the charge of the ion. Collisional 
recombinations and de-excitations follow from the principle of 
detailed balance. Detailed calculations should replace these approximate 
values where available (see, e.g., Nahar \& Pradhan 1997 and Nahar 1999).

\section{Summary}

We have smoothed the photoionization cross-sections of the Opacity
Project for neutral  Na and S,  singly ionized Ne,   and  neutral and
singly-ionized  Li, Be, B, C, N, O, F, Mg, Al, Si and Ca, using a
Gaussian profile with $\sigma \equiv \delta E = 0.03 E$, where $E$ is
the energy of the ionizing photon. This procedure  follows  Bautista et
al. (1998). The smoothed cross-sections have been merged  with the
energy levels and line transition probabilities listed in the Atomic
and Spectroscopic Database at NIST to derive model atoms and ions
suitable for non-LTE calculations of model atmospheres and line
formation. This new  collection of model atoms complements and, for the
species in common, supersedes a previous dataset,  the {\sc modion}
model atoms, developed for early-type stellar atmospheres.

As an example, we show the results of two calculations involving
the RAP models. 
Figure 3 shows the relative importance of neutral C, N, O, and Mg on the
continuum absorption in the near-UV spectrum of the Sun ($T_{\rm eff} =
5777$ K; $\log g = 4.44$ dex; [Fe/H]=0.0 dex) and Procyon  ($T_{\rm
eff} = 6530 $ K; $\log g = 4.0$ dex; [Fe/H]=0.0 dex). The spectral
synthesis assumes LTE
 and makes use of  Kurucz's atmospheric structures (Kurucz 1993).

Although the models are based on observed energy levels and lines,
 together with  detailed calculations  of photoionization
 cross-sections, judging their appropriateness for non-LTE
 calculations in stellar atmospheres is  beyond the scope of this
 paper.  
 Several approximations and simplifications are
 implicit in the models and, therefore, caution is required. Only  
detailed  comparison with high quality observed spectra will determine
 the models' usefulness.    More complex ions not yet included in
 TOPBASE are of interest. Iron is at the head of this group, and new
 calculations of photoionization cross-sections are available (Bautista
 1997) -- but not as part of TOPBASE at the time of this writing. 
 Many other sources of data, although with a significant
 heterogeneity,  exist and should be used to complement,  check,  and
 improve these models. Such tasks and a detailed comparison with
 high-quality observations  for a series of  standard stars are in progress
 and will be reported in the  future.

Both, the {\sc modion} models\footnote{\tt http://tlusty.gsfc.nasa.gov} and
the RAP models\footnote{\tt http://hebe.as.utexas.edu/at/at.cgi}
(and, independently, the RAP photoionization cross-sections) 
are publicly available.

\acknowledgments

 We thank the people and institutions involved in creating and
 maintaining the ASD/NIST and TOPBASE databases, which have been
 extensively used in this work. This research has been supported by NSF
 through grant AST-0086321 and by the Robert A. Welch Foundation of
 Houston, Texas.

\clearpage

\begin{deluxetable}{crrcrr}
\tablecaption{Number of Levels and Radiative Transitions Included in the Different Models. \label{table1}}
\tablewidth{0pt}
\tablehead{
\colhead{Model} & \colhead{Num. Levels}   & \colhead{Num. Transitions}   & \colhead{Model} & \colhead{Num.  Levels}   & \colhead{Num. Transitions}}
\startdata
\ion{Li}{1} & 26 & 153 & \ion{Li}{2} & 37 & 143 \\
\ion{Be}{1} & 56 & 323 & \ion{Be}{2} & 18 & 72 \\
\ion{B}{1} & 29 & 117 & \ion{B}{2} & 38 & 134 \\
\ion{C}{1} & 104 & 1090 & \ion{C}{2} & 40 & 189 \\
\ion{N}{1} & 89 & 615 & \ion{N}{2} & 51 & 280 \\
\ion{O}{1} & 54 & 242 & \ion{O}{2} & 74 & 512 \\
\ion{F}{1} & 52 & 209 & \ion{F}{2} & 73 & 326 \\
\dots & \dots & \dots & \ion{Ne}{2} & 59 & 353 \\
\ion{Na}{1} & 32 & 192 & \dots & \dots & \dots \\
\ion{Mg}{1} & 71 & 434 & \ion{Mg}{2} & 31 & 184 \\
\ion{Al}{1} & 33 & 160 & \ion{Al}{2} & 81 & 537 \\
\ion{Si}{1} & 57 & 215 & \ion{Si}{2} & 46 & 261 \\
\ion{S}{1} & 64 & 313 & \dots & \dots & \dots \\
\ion{Ca}{1} & 79 & 543 & \ion{Ca}{2} & 32 & 147 \\
  \enddata
\end{deluxetable}


\begin{thebibliography}{}
\bibitem[Allende Prieto, Garc\'{\i}a L\'opez, Lambert \& Ruiz 
Cobo (2000)]{2000ApJ...528..885A} Allende Prieto, C. , Garc\'{\i}a L\'opez, R. J., Lambert, D. L. \& Ruiz Cobo, B.  2000, \apj, 
528, 885 


\bibitem[Asplund et al.(2000)]{2000A&A...359..729A} Asplund, M., Nordlund, 
{\AA}., Trampedach, R., Allende Prieto, C., \& Stein, R.~F.\ 2000a, \aap, 
359, 729. 

\bibitem[]{} Auer, L H. \& Mihalas, D. 1973, \apj, 184, 151


\bibitem[Bautista (1998)]{} Bautista,  M. A. 1997, \aaps, 122, 167

\bibitem[Bautista, Romano \& Pradhan (1998)]{1998ApJS..118..259B} Bautista,  M. A., Romano, P.  \& Pradhan, A. K. 1998, \apjs, 118, 259 

\bibitem[Brage, Judge \& Brekke (1996)]{1996ApJ...464.1030B} Brage, T. , 
Judge, P. G. \& Brekke, P. 1996, \apj, 464, 1030 

\bibitem[Butler (1998)]{1998amda.conf...23B} Butler, K. 1998, Atomic and 
Molecular Data and their Applications, AIP Conference Proceedings, Peter J. Mohr and Wolfgang L. Wiese, eds., vol. 434, 1998., p.23 


\bibitem[Cunto, Mendoza, Ochsenbein \& Zeippen (1993)]{1993A&A...275L...5C} 
Cunto, W., Mendoza, C., Ochsenbein, F. \& Zeippen, C. J. 1993, \aap, 275, L5 

\bibitem[Dere, et al. (1997)]{1997A&AS..125..149D} Dere, K. P., Landi, E., 
Mason, H. E., Monsignori Fossi, B. C. \& Young, P. R. 1997, \aaps, 125, 149 


\bibitem[Hauschildt, Allard \& Baron (1999)]{1999ApJ...512..377H} 
Hauschildt, P. H., Allard, F.  \& Baron, E. 1999a, \apj, 512, 377 

\bibitem[Hauschildt, et al. (1999)]{1999ApJ...525..871H} Hauschildt, P. H., 
Allard, F. , Ferguson, J. , Baron, E. \& Alexander, D. R. 1999b, \apj, 525, 
871 


\bibitem[Hubeny (1988)]{} Hubeny, I. 1988, Computer Phys. Comm., 52, 103

\bibitem[Hubeny \& Lanz (1995)]{1995ApJ...439..875H} Hubeny, I. \& Lanz, T. 
1995, \apj, 439, 875 

\bibitem[Hubeny \& Lanz (1997)]{} Hubeny, I. \& Lanz, T., 1997, TLUSTY  User's Guide, Version 195 ({\tt http://tlusty.gsfc.nasa.gov/})


\bibitem[Kupka, et al. (1999)]{1999A&AS..138..119K} Kupka, F., Piskunov, 
N., Ryabchikova, T. A., Stempels, H. C. \& Weiss, W. W. 1999, \aaps, 138, 
119 

\bibitem[Kurucz(1993)]{1993KurCD..13.....K} Kurucz, R.\ 1993, ATLAS9 
Stellar Atmosphere Programs and 2 km/s grid.~Kurucz CD-ROM No.~13.~ 
(Cambridge, Mass.: Smithsonian Astrophysical Observatory). 


\bibitem[Lanz et al. (1996)]{} Lanz, T., Hubeny, I. \& de Koter, A. 1996, Phys. Scr., T65, 144

\bibitem[Martin (1997)]{} Martin, W. C., 1997, Working Group 1 Report for the IAU, Commission on Atomic and Molecular data ({\tt http://physics.nist.gov/PhysRefData/datarefs/contents.html})

\bibitem[]{} Mihalas, D.  1972, \apj, 177, 115

\bibitem[]{} Mihalas, D.  1978, Stellar Atmospheres, 2$^{\rm nd}$ Edition, (San Francisco: Freeman), p. 133

\bibitem[]{} Mihalas, D., Heasley, J. N., \& Auer, L. H. 1975, NCAR Technical Note NCAR-TN/STR-104

\bibitem[]{} Nahar, S. N. 1999, \apjs, 120, 131

\bibitem[]{} Nahar, S. N.,\& Pradhan, A. K. 1997, \apjs, 111, 339


\bibitem[]{} Plez, B., Brett, J. M., \& Nordlund, \AA. 1992, A\&A, 256, 551

\bibitem[]{} Pradhan, A. K. \& Peng, J. 1995, in Analysis of Emission Lines, R. E. Williams and M. Livio, eds., STScI Symposium Series No. 8, (Cambridge: Cambridge University Press) 

\bibitem[]{} Pradhan, A. K. \& Zhang, H. L. 2000, Landolt-B\"ornstein: 
Numerical Data and Functional    Relationships in Science and Technology,
   Y. Itikawa, ed., (Berlin: Springer),  in press


\bibitem[]{} Seaton, M. J. 1962, in Atomic and Molecular Processes, ed. D. R. Bates (New York: Academic Press), p. 374

\bibitem[Stein \& Nordlund (1998)]{1998ApJ...499..914S} Stein, R. F. \& 
Nordlund, \AA. 1998, \apj, 499, 914 


\bibitem[]{} Tsuji, T. 2002, \apj, 575, 264

\bibitem[]{} Van Regemorter, H. 1962, \apj, 136, 906

\bibitem[]{} Varosi, F., Lanz, T., deKoter, A., Hubeny, I., \& Heap S.R. 1995, 
{\tt ftp://idlastro.gsfc.nasa.gov/pub/contrib/varosi/modion}


\bibitem[Verner, Ferland, Korista \& Yakovlev (1996)]{1996ApJ...465..487V} 
Verner, D. A., Ferland, G. J., Korista, K. T. \& Yakovlev, D. G. 1996, 
\apj, 465, 487 

\bibitem[Wiese (1997)]{} Wiese, W. L., 1997, Working Group 2 Report for the IAU, Commission on Atomic and Molecular data ({\tt http://physics.nist.gov/PhysRefData/datarefs/contents.html})


\end{thebibliography}
\end{document}